# HoneyMesh: Preventing Distributed Denial of Service Attacks using Virtualized Honeypots

Hrishikesh Arun Deshpande
Member of Technical Staff R&D,
NetApp India Pvt. Ltd,
Bangalore, India

*Abstract*— Today, internet and web services have become an inseparable part of our lives. Hence, ensuring continuous availability of service has become imperative to the success of any organization. But these services are often hampered by constant threats from myriad types of attacks. One such attack is called distributed denial of service attack that results in issues ranging from temporary slowdown of servers to complete non-availability of service. Honeypot, which is a sort of a trap, can be used to interact with potential attackers to deflect, detect or prevent such attacks and ensure continuous availability of service. This paper gives insights into the problems posed by distributed denial of service attacks, existing solutions that use honeypots and how a mesh of virtualized honeypots can be used to prevent distributed denial of service attacks.

## I. INTRODUCTION

In today's world of technology and computers, internet serves as a critical platform for both service providers and consumers. The success of any venture is critically dependent on reliability and continuous availability of service. Thus, it's crucial for service providers to protect their servers from various security threats and attacks. Of all the attacks that hinder the availability of service, a denial of service attack poses maximum threat to an organization since it has direct effect on the service availability to a consumer. A denial of service attack results in a temporary or long-term non-availability of a service to its intended users by the way of either crashing a service resulting in complete non-availability or by flooding a server with fraudulent requests thereby slowing down the delivery of service to real users [1]. Honeypot can be used as an intrusion detection mechanism that can replicate some or all actions of a server and effectively monitor potential attackers thereby enabling the server admins to detect and prevent potential denial of service attacks to ensure a reliable and continuous service to their intended users.

## II. DISTRIBUTED DENIAL OF SERVICE ATTACKS

A denial of service (DoS) attack is an attempt to make a service, usually one offered over internet, unavailable to its legitimate users [1-3]. This can result in either temporary interruption in service by means of overwhelming the server with several requests or a permanent one that causes the server to crash.

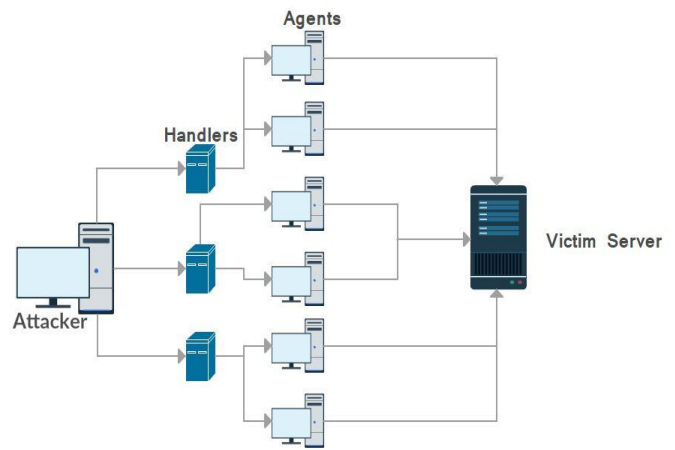

Fig. 1 Distributed Denial of Service Attack (DDoS)

A more hazardous version of DoS is a distributed denial of service (DDoS) attack. A distributed denial of service attack is the one in which intruders execute attacks from multiple locations rather than a single one [2]. An attacker initially compromises and gains control over local systems. These compromised systems are called handlers or masters. Masters are then used to further compromise systems that are close to the target server. These systems called as DDoS agents or slaves are then used to launch multiple attacks on the victim server [3]. In most cases, IP addresses of attack sources are forged to make the identification of the attacker location difficult. A combined effect of relaying the attack from several DDoS agents can potentially overwhelm the server with thousands of requests resulting in a slowdown or crash.

### A. Types of Distributed Denial of Service Attacks

DDoS attacks are of different types depending on the kind of attack and the effect they have on the target server. As such several DDoS attacks have been identified till date. Based on the impact they have on the target server, DDoS attacks can be broadly classified into two major types namely flood attack and crash attack [2-3]. Flood attack involves overwhelming the target server with several thousand requests thereby slowing down the service for legitimate users [4]. On the other hand, a crash attack exploits the vulnerabilities in the victim server causing it to hang, reboot or crash. Table I summarizes various types of DDoS attacks and their effects.

TABLE I. Types of DDOS attacks

| Ser. No | Attack | Effect |
|---|---|---|
| 1 | Smurf attack | Forged ICMP packets are sent to the destination server which responds with ICMP reply packets thereby flooding the server with fake requests and denying service to real users. |
| 2. | TCP/SYN Flood | The target server is sent TCP packets with unreachable addresses. The server wastes all its time and resources in determining the right destination causing denial of service to others. |
| 3. | UDP Flood Attack | This happens when the attacker sends a forged UDP packet to a port which responds with a destination unreachable ICMP response. This floods the system if several UDP packets are sent. |
| 4. | Teardrop | Here, jumbled overlapping TCP/IP fragments are sent to the victim server which can crash the system due to difficulty in reassembling the overlapping fragments. |
| 5. | Ping of Death | In this case, the destination server is sent an ICMP packet much larger than its expected size. The victim server is unable to reassemble the packet and crashes as a result. |
| 6. | Land Attack | This happens when an attacker sends a packet with identical source and destination addresses. This confuses the target server resulting in a crash. |
| 7. | Ping Flood | This is the most common of the DDOS attacks. Here the attacker sends repeated ping commands to a server resulting in flooding. |
| 8. | Nuke attack | The destination server is flooded with counterfeit ICMP packets that exploit the vulnerabilities of Operating systems causing the system to halt. |

*B. Effects of Distributed Denial of Service Attacks*

DDOS attacks are known to disrupt services causing inconvenience to intended users. The effects of such attacks can be either temporary or permanent [3]. Typical temporary effects include flooding, slowdown of services, rapid consumption of resources, sudden spikes in processor usage etc. These attacks manifest themselves as temporary outages that cause non-availability of service for short periods of time [4]. On the other hand, permanent attacks are catastrophic and can result in server crashes, disruption of routing information, data corruption or in extreme cases render the server hardware unusable requiring complete hardware replacement. Such attacks can result in long term outages and can severely damage the reputation of an enterprise resulting in a decline of user trust. Thus it has become imperative to protect vulnerable servers against such attacks [1-4].

## III. HONEYPOTS

Since distributed denial of service attacks can be potentially harmful to a target server, it's essential to effectively detect and reduce such attacks. Although absolute prevention of attacks is difficult, several techniques have been proposed to counter DDoS attacks. The two main techniques that deal with DDoS attacks involve mitigation of attacks and identification of the attack source [4]. Honeypots can be effectively used in both of these cases. Fig.2 illustrates the design of a basic honeypot.

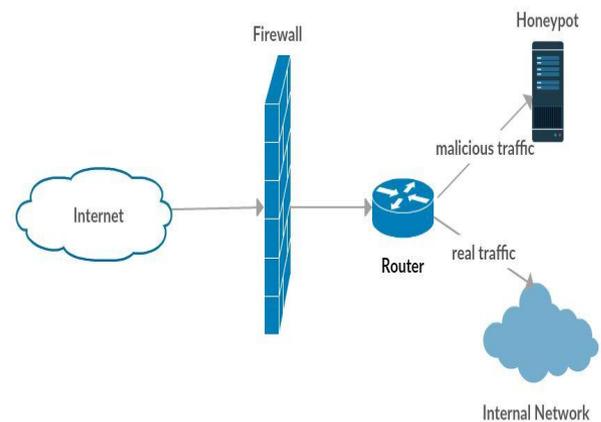

Fig. 2 Basic Honeypot design

There are several ways in which a honeypot can be defined. In simplest terms, a honeypot can be defined as a trap for an attacker that mimics some or all activities of a real system and records the activities of the attack source [5]. Honeypots can be used in a flexible manner at the server side to not only detect such attacks but to also protect the user's critical data and record possible malicious activities so as to track the attacker. Honeypots can be broadly classified into two categories namely low interaction and high interaction honeypots [6]. High interaction honeypots imitate most services of real production systems and host a variety of tasks. They provide more security and are hard to detect but are relatively expensive to maintain. On the other hand, low interaction honeypots simulate services that are frequently requested by attackers. They consume fewer resources and can be easily maintained [7]. Both types of honeypots can be implemented as virtual machines and hosted on a single physical server [8].

*A. Existing Solutions*

Due to the potential threats posed by DDoS attacks, several solutions have been already proposed to deal with these threats. Bellovin's ITRACE uses ICMP packets to determine the path of a small subset of forwarded packets enabling the victim to identify a compromised DDoS agent [9]. Other solutions include marking the path traversed by packets to determine their real source and reduce the number of markings by utilizing the topology of network maps. But these solutions are inefficient owing to the fact that they assume a large part of the network to implement them and thus they fail to address specific threats posed by DDoS attacks at production servers.

Some solutions also propose the usage of honeypots to mitigate DDoS attacks. Weiler proposes implementation of a cluster of physical honeypots servers that mimic the activities of real servers [10]. This solution is expensive since every honeypot needs a separate physical server which results in wastage of resources and high maintenance costs. Das proposes to mitigate denial of service attacks using a concept called "Active Servers" (AS) [11]. Every production server is hidden behind an AS that acts like a gateway to the production server. Legitimate traffic is passed on to real server while malicious traffic is halted. For malicious traffic, an AS acts like a honeypot thus protecting real server from being compromised. This solution is robust and secure but it slows down processing of requests for real users since each and every request needs to pass through an additional gateway. Moreover, separate honeypot servers for each production server wastes resources and is quite expensive. Also flooding attacks with thousands of requests can clog these gateway servers thereby greatly slowing down the access to production servers. This actually ends up aiding the attacker by slowing down the service for real users. Khattab proposed another solution to mitigate denial of service attacks where honeypots and production servers are frequently shuffled within the network [12]. Honeypots are used to detect and prevent DoS attacks. This solution is effective when most incoming requests are DoS requests. But if majority of traffic is legitimate and only few requests are DoS attacks, the solution is ineffective since a certain number of servers function as honeypots irrespective of the traffic. This again wastes resources and constant shuffling of honeypots and production servers in fact slows down service for intended users. Sridhar proposes the usage of honeypots to prevent DDoS attacks for cloud infrastructure [13]. This solution proposes a network of honeypots to monitor attacker activities but doesn't provide satisfactory solutions to mitigate flooding attacks.

## IV HONEYMESH- A NETWORK OF VIRTUALIZED HONEYPOTS TO PREVENT DDOS ATTACKS

The proposed solution is to create a network of virtualized honeypots within the existing infrastructure with minimal cost and maintenance overheads. The existing security infrastructure consists of services such as ftp, mail, web and DNS that are offered to the outside world through a demilitarized zone (DMZ) [14]. DMZ consists of two firewalls. The first firewall is meant to protect these servers from external malicious traffic while the second one is an internal firewall meant to protect the organization's internal network. The two firewall approach provides multiple layers of protection to the internal network. In addition to this, other security mechanisms such as encryption, host based intrusion detection systems, vulnerability scanners are used to bolster protection. Further, the organization might choose to add further protection to its local services using a virtual private network (VPN). These mechanisms contribute towards securing the network. However, effective detection and deflection of attacks together with identification of attack sources is necessary. This is accomplished using honeypots.

Unlike earlier solutions that used explicit servers as honeypots to function as gateways and mimic a real server, the new solution proposes to implement honeypots as virtual machines (VM) that can be hosted on a few physical servers [15]. Since VM's share resources, multiple honeypots can be hosted on a single server [16] as shown in Fig. 3.

Honey VM's have security mechanisms similar to the real servers but some vulnerabilities are deliberately exposed so as to lure the attacker into a trap [5][10][13]. These VM's continuously monitor the incoming traffic for potential malicious activities and once an attack is discovered, all the traffic from the attack source is routed to the honey VM network. This ensures that malicious traffic doesn't reach the production servers. Also, each honeypot can be customized to mimic specific servers. For example, one honey VM can mimic a file server while another can imitate a web server. This forms a network of virtual honeypot servers that constitutes a honeypot farm.

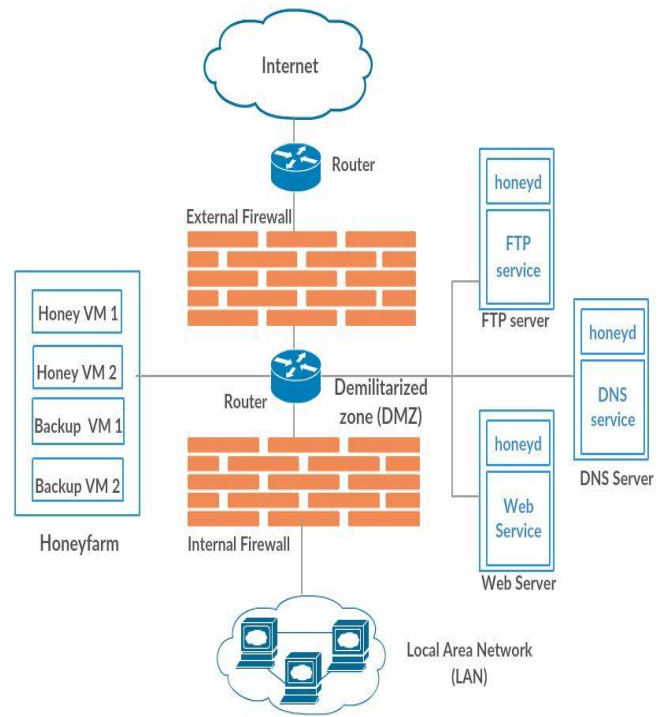

Fig. 3 Honeymesh security infrastructure

Additionally, each of these honeypots can have backup VM's that normally remain idle but can be activated the moment an existing honey VM is compromised by an attacker. This ensures that intrusion detection and deflection is not halted when an existing honeypot is compromised by a DDoS attack. This arrangement functions like a hybrid honeypot network that mimics the functionalities of real servers similar to high interaction honeypots while consuming fewer resources just like low interaction honeypots [16].

As opposed to the solution proposed by Das where separate honeypot servers' function as gateways to individual production servers [11], the gateway honeypot can run as a daemon process within the server itself. This honeypot daemon, abbreviated as honey-d, works like a gateway and performs initial authentication before passing on the information to the actual server. Thus even if the honey farm fails to detect an attack, honey daemon present within the server provides an additional layer of security. This, together with the hybrid network of honey VM's functions like a mesh of virtualized honeypots and ensures effective detection and prevention of possible DDoS attacks.

*A. Detection of an Attack*

Honeypot VM's in the honey farm employ machine learning algorithms to perform a behavioral analysis of incoming traffic [18]. Since each production server receives different types of requests, appropriate honey VM's can be tailored for the corresponding servers. For example, one honey VM can analyze web server traffic while another can examine file server requests. After analyzing a few thousand requests, each honey VM generates a baseline model of expected traffic. Incoming requests are compared against the baseline. If any deviation is observed, further probing is necessary to confirm if the request actually constitutes a DDoS attack.

Once the honeypot suspects a particular request based on behavioral analysis, it needs to verify that the suspicious request is actually a DDoS attack. For this, the honey VM then employs a challenge-response model to gather more information. This is accomplished by sending a set of challenge queries to suspicious source [19]. Based on the responses received, the honey VM decides whether further investigation is necessary. If yes, more sophisticated challenges are sent to the source. Based on the responses received and an intelligent behavioral mechanism, the honey VM can conclude whether the requests constitute a DDoS attack. This process is fully automated and happens without human intervention thereby guaranteeing excellent service for legitimate users.

Similar mechanisms can be built into the honey daemons that run on production servers. This ensures that even if the honey farm misses out on a potential attack, it is reexamined by honey-d's running on respective servers. This provides an additional level of authentication and intrusion detection.

### B. Preventing Flooding Attacks

Once an attack is discovered, the routing information in the internal routers is modified so as to redirect all incoming traffic from the attack source to the honey farm. Since malicious traffic now flows to the honey farm, it ensures that the production network is shielded from flooding attacks. Honey VM's in the farm keep the attacker engaged through a set of challenge-response queries further slowing down the attacker [19]. Also, once the attack source is confirmed, all incoming traffic from that source is blocked at the firewall itself. This mechanism mitigates the impact of flooding attacks to a great extent.

### C. Preventing Crashing Attacks

Unlike flooding attacks that cause short term outages and slowdown of services, crashing attacks manipulate vulnerabilities in production servers causing data corruption, theft of confidential information, server crashes and reboots causing long term outages [4]. Thus, preventing such attacks requires additional intelligence in honey VM's and daemons.

For this purpose, each VM in the honey farm is made to mimic most services of real servers and implement most security mechanisms provided for real servers. This lures the attacker into believing that interactions are happening with real servers [5][6][10][13]. The VM then sets a trap for the attacker by deliberately exposing some flaws in its security mechanism to fool the attacker into thinking that a DDoS attack has succeeded. Meanwhile, the VM tracks the attack source through a challenge-response mechanism and further requests from the attack source are blocked by the firewall.

Since a honey VM exposes some security flaws to the attacker, there's a chance that the VM can be compromised or crash in worst case scenarios. The new design proposes to maintain backup VM's that can immediately take charge if the current VM is compromised. By maintaining a pool of backup VM's, we can ensure continued intrusion detection and prevention. Also since the compromised honeypot is a VM, it can be easily restored with minimal cost [17].

Fig. 4 illustrates a typical sequence of events that occur in the honey mesh while detecting and preventing DDoS attacks.

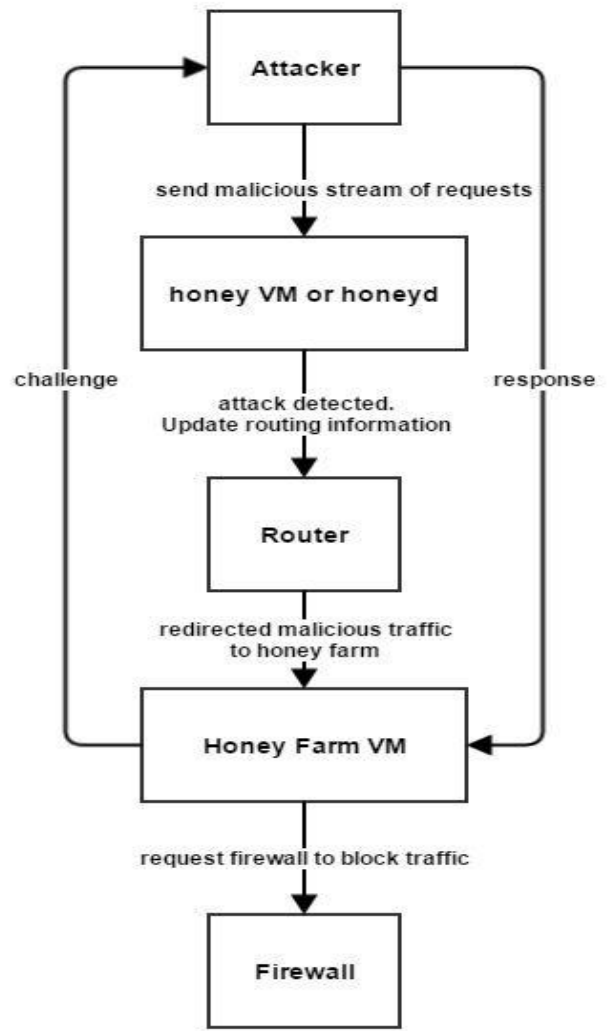

Fig. 4 Honeymesh sequence flow

### D. Advantages and Future Enhancements

The proposed solution of preventing DDoS attacks by creating a mesh of honey VM's and honey daemons has several advantages over the existing solutions. These advantages include:

- Since honeypots are implemented as virtual machines and daemon processes rather than actual physical servers, this solution is economical and has low maintenance costs [17].
- Also, each honey VM is backed up by additional VM's. This ensures continuous intrusion detection and prevention even if an existing VM is compromised.
- Restoring a compromised VM is very cheap and has minimum downtime [17].
- The mesh of honey VM's and daemons provides multiple layers of security against DDoS attacks. Even if honey VM's in the honey farm miss out a possible attack, it can be caught by honey-d's running on individual servers. This authentication provides enhanced security to the production servers.

- Since malicious traffic is routed to the honey farm, actual production servers and network lines are protected from flooding.
- Challenge response and behavioral analysis by honeypots ensures effective intrusion detection and prevention of crashing attacks [18-19].

Despite all the advantages mentioned, the proposed solution has a few shortcomings which have been stated as below:

- Although production servers and organization's internal network (LAN) are fully protected, there should be a mechanism to protect the organization's routers from being flooded with malicious requests.
- Honeypot VM's may be hosted on a network of servers to create a more robust honey farm. Currently, all VM's in the honey farm are hosted on a single server to reduce maintenance and recovery costs.
- Honeywalls where the honeypot logic is embedded within the firewall itself can be implemented.

Although the solution has a few shortcomings as mentioned above, honeymesh is very robust and provides multiple levels of security checks and intrusion detection mechanisms to effectively detect deflect and prevent possible DDoS attacks. Also, the above shortcomings can be addressed in future enhancements of the proposed solution.

## V CONCLUSION

Distributed denial of service (DDoS) attacks are dangerous and can potentially render the production site unusable either by flooding the server network with thousands of malicious requests or crashing the server by exploiting the vulnerabilities in its software. Several solutions have been proposed to deal with DDoS attacks. However, these solutions are either expensive due to usage of multiple physical servers for honeypots or do not successfully address the issue of flooding type of DDoS attacks. The new solution proposes to create a virtual network or mesh of honeypot VM's and honey daemon processes to provide multiple levels of security checks and intrusion detection using behavioral analysis and challenge-response models. Also, malicious traffic is routed to honey farm thereby protecting the production server and internal networks from both crashing and flooding type of DDoS attacks. Honeymesh when integrated with existing security infrastructure such as firewalls, encryption, authentication services, virtual private network (VPN) etc. can protect the server network from any kind of DDoS attacks. As already stated, the solution does have a few shortcomings which can be addressed in future enhancements to this solution.